\documentclass[aps,twocolumn,showpacs]{revtex4}
\usepackage{graphicx}
\usepackage{epsfig}

\begin{document}

\title{Driving-dependent damping of Rabi oscillations in two-level semiconductor systems}

\author{D. Mogilevtsev$^{1}$, A. P. Nisovtsev$^{1}$,  S. Kilin$^1$, S. B.
Cavalcanti$^{2,3}$, H. S. Brandi$^{4,5}$ and L. E. Oliveira$^{3}$ }

\affiliation{$^1$Institute of Physics, NASB, F. Skarina Ave. 68, Minsk, 220072, Belarus \\
$^2$Departamento de F\'{\i}sica, UFAL,
Cidade Universit\'{a}ria, 57072-970, Macei\'{o}-AL, Brazil \\
$^3$Instituto de F\'{i}sica, UNICAMP, CP 6165, Campinas - SP, 13083-970, Brazil \\
$^4$Instituto de F\'{i}sica, UFRJ, Rio
de Janeiro-RJ, 21945-9702, Brazil \\
$^5$ Inmetro, Campus de Xerem, Duque de Caxias-RJ, 25250-020, Brazil}

\begin{abstract}
We propose a mechanism to explain the nature of the damping of Rabi
oscillations with increasing driving-pulse area in localized
semiconductor systems, and have suggested a general approach which
describes a coherently driven two-level system interacting with a
dephasing reservoir. Present calculations show that the
non-Markovian character of the reservoir leads to the dependence of
the dephasing rate on the driving-field intensity, as observed
experimentally. Moreover, we have shown that the damping of Rabi
oscillations might occur as a result of different dephasing
mechanisms for both stationary and non-stationary effects due to
coupling to the environment. Present calculated results are found in
quite good agreement with available experimental measurements.

\end{abstract}

\maketitle

Localized semiconductor systems exhibiting few discrete energy
levels ("artificial atoms"), such as specially selected donor
impurities and quantum dots (QDs), are prospective candidates to
play the role of basic building blocks for quantum information
processing. In particular, a two-level semiconductor system may
exhibit Rabi oscillations (ROs) of its population when coupled to
a driving field, so that it may be coherently controlled
\cite{experiment,exper,no-biexciton,int_indep,traps}. There are a
number of dephasing mechanisms for localized semiconductor
systems, some of which are essentially non-Markovian so that one
needs to take into account memory effects as well as the
back-action of a dissipative reservoir on the radiating system.
For example, a dephasing caused by spin-spin coupling between
neighboring QDs or carriers captured in traps in the vicinity of a
QD was shown to lead to non-Markovian dynamics
\cite{apn-works,spin}. Such reservoirs have correlation times
comparable with the typical decoherence time of the dephasing
system. Also, the dephasing due to coupling with phonons was shown
to lead to non-Markovian features in the dynamics of a two-level
systems (TLS) \cite{phonon}. Carriers and excitons in localized
semiconductor systems may be coupled not only to localized
neighboring  states, but also to delocalized ones \cite{deloc}.
This diversity of dissipation channels has led to a number of
novel features in such systems' dynamics. In the present work we
focus our attention on one peculiar phenomenon which has caused
and is still causing much controversy, namely, the damping of ROs
due to the increase of the driving-pulse area which is an observed
feature of coherently excited localized semiconductor systems
\cite{experiment,exper,no-biexciton,int_indep,traps}. A number of
mutually contradicting explanations was suggested for it. One of
these is that such a dephasing is due to the system's interaction
with a non-Markovian reservoir of phonons \cite{phonon}. However,
the dephasing process takes place even when the coupling with
phonons is negligible \cite{no-biexciton}. Driving-dependent
damping of ROs was proposed to occur as a consequence of
excitations of bi-excitons in the QD \cite{biexciton}, although
damped ROs are also observed when there is no possibility for the
bi-exciton excitation \cite{no-biexciton}. Recently, it was
demonstrated that the experimentally observed \cite{exper}
intensity-dependent damping of ROs can be reproduced by
introducing into the standard Bloch equations a dephasing rate
dependent on the driving-field intensity \cite{oliveira}.  On the
other hand, although there is an experimental confirmation of a
driving dependence of the dephasing rate \cite{no-biexciton}, an
intensity-independent dephasing rate has also been measured
\cite{int_indep}.

Based on this controverted scenario, in the present work we propose
to shed some light on this matter by studying a simple TLS excited
by a classical coherent field and coupled to a general dephasing
reservoir. Within a quite general and straightforward approach, we
demonstrate that a driving-field dependent damping of ROs stems from
various relaxation mechanisms entering into play in different
experimental situations. Furthermore, we show that driving-dependent
damping may occur whether the reservoir is influenced or not by the
driving field. To keep it simple, and to focus only on features which give rise to
the phenomenon in question, we assume \textit{no} population damping
of the TLS. This also corresponds to the real experimental situation
with driving by short laser pulses, so that the population damping
is negligibly slow on the time-scale of the system's dynamics
\cite{exper}. In the frame rotating with the driving-field
frequency, $\omega_L$, working within the interaction picture with
respect to the reservoir variables and using the rotating-wave
approximation (RWA), we describe our problem with the following
standard effective Hamiltonian,
\begin{equation}
H_{tot}(t)=H_0(t)+\hbar\sigma^+\sigma^-{\bf R}(t), \label{ham1}
\end{equation}
where the undamped system's Hamiltonian is given by
\begin{equation}
H_{0}(t)=\hbar\Delta\sigma^+\sigma^-+ \hbar
[\Omega(t)\sigma^++\Omega^*(t)\sigma^-]. \label{ham0}
\end{equation}
Here $\sigma^{\pm}=|\pm\rangle\langle\mp|$ are the system's raising
and lowering operators, the kets $|\pm\rangle$ correspond to the
excited and ground states of the TLS, respectively,
$\Delta=\omega_0-\omega_L$ is the detuning of the driving-laser
frequency $\omega_L$ from the resonance frequency $\omega_0$ of the
TLS transition, with the possible addition of a frequency-shift term
due to the interaction with the dephasing reservoir. The reservoir
is described by the operator ${\bf R}(t)$, which might also depend
on classical stochastic variables (describing, for example,
different realizations of the reservoir in each run of an experiment
\cite{apn-works}), whereas $\Omega(t)$ describes the shape of the
driving pulse.

Let us now assume that the reservoir correlation function
$\langle{\bf R}(t){\bf R}(\tau)\rangle$ satisfies the following
general requirements: $\langle{\bf R}(t){\bf
R}(\tau)\rangle\rightarrow 0$, when $t,\tau\rightarrow \infty$,
and $|t-\tau|\rightarrow \infty$. If the coupling of the reservoir
to the TLS is weak, and the reservoir correlation function decays
with $t,\tau\rightarrow \infty$, and also with
$|t-\tau|\rightarrow \infty$ much faster than the typical
time-scale of the system's evolution, it is possible to obtain a
time-local master equation \cite{{apn-works},{tlmast}} for the
problem described by the Hamiltonian (\ref{ham1})-(\ref{ham0}).
Following the approach developed in Refs. \cite{apn-works},  we
introduce dressed operators describing the interaction with a
classical field, i.e., ${\bf
S}^{\pm}(t)=U^{\dagger}(t)\sigma^{\pm}U(t)$, where the unitary
"dressing" transformation \cite{dress} is given by:
$U(t)=\overleftarrow{{\bf
T}}\exp\left\{-{i\over\hbar}\int\limits_{t_0}^tH_0(\tau)d\tau\right\}$,
and $\overleftarrow{{\bf T}}$ denotes the time-ordering operator.
One may use the time-convolutionless projection operator technique
or cumulant's expansion and the Born approximation for the
"dressed" density-matrix master equation, and then going back to
the "bare" basis, one obtains the following set of Bloch equations
with time-dependent coefficients \cite{{apn-works},{tlmast}}:
\begin{eqnarray}
\frac{d\rho_{++}}{dt}&=&i\left[\Omega(t)\rho_{-+}-\Omega^*(t)\rho_{+-}\right],
\label{bloch1}
\\
\frac{d\rho_{+-}}{dt}&=&\left\{i[\Delta+\langle{\bf
R}(t)\rangle]-\kappa(t)\right\}\rho_{+-} \label{bloch2} \\
\nonumber &+& i\overline{\Omega}^*(t)[1-2\rho_{++}],
\end{eqnarray}
where $\rho_{++}=\langle+|\rho|+\rangle$,
$\rho_{\pm\mp}=\langle\pm|\rho|\mp\rangle$, $\rho$ is the density
matrix of the TLS in the "bare" basis, and the time-dependent
dephasing rate $\kappa(t)$ and the generalized Rabi frequency
$\overline{\Omega}(t)$ are defined as
\begin{eqnarray}
\kappa(t)&=&\int\limits_{t_0}^t d\tau \langle{\bf R}(\tau){\bf
R}(t)\rangle D_{++}(\tau-t), \label{kappa1} \\
\overline{\Omega}(t)&=&{\Omega}(t)-\int\limits_{t_0}^t d\tau
\langle{\bf R}(\tau){\bf R}(t)\rangle D_{+-}(\tau-t) \, ,
\label{kappa2}
\end{eqnarray}
where $D_{+-}(t)$ and $D_{++}(t)$ are dressing functions
\cite{dress}. In the case of a rectangular pulse ($\Omega(t)\equiv
\Omega/2$ for the pulse duration, which we will use in further
discussions here), one obtains
\begin{eqnarray}
D_{++}(t)&=&\frac{1+c^2+s^2\cos(\Omega_{R}t)}{2}, \label{dfunctions1} \\
D_{+-}(t)&=&\frac{i\Omega}{\Omega_{R}} \left\{c[1-\cos(\Omega_{R}t)]
+i\sin(\Omega_{R}t)\right\}, \label{dfunctions2}
\end{eqnarray}
where $c={\Delta/\Omega_{R}}$, $s={\Omega/ \Omega_{R}}$, and
$\Omega_{R}=\sqrt{\Delta^2+\Omega^2}$ is the effective Rabi
frequency.

Let us now consider the simplest situation in which the driving
field interacts only with the localized system. In the Markovian
limit one has $\langle{\bf R}(\tau){\bf R}(t)\rangle\sim
\delta(\tau-t)$ and, therefore, as follows from Eqs.
(\ref{kappa1})-(\ref{kappa2}), one recovers the standard system of
Bloch equations for a driven TLS in the presence of dephasing
effects. In this case, the  dephasing rate, $\kappa(t)\equiv\kappa$, is
constant and independent of the driving-field intensity.  Then, as
expected, ROs persist for all values of the
field's intensity [cf. dotted curve in Fig. 1(a)].

For a general non-Markovian reservoir one may write the
reservoir's correlation function as a sum of a stationary
contribution $K(\tau-t)$ and a non-stationary one $P(\tau,t)$
which tends to zero for $t=\tau$ as $t,\tau\rightarrow \infty$,
i.e., $ \langle{\bf R}(\tau){\bf
R}(t)\rangle=K(\tau-t)+P(\tau,t)$, where $P(\tau,t)$ is
responsible for non-Markovian effects at the initial stage of the
system's dynamics. For the moment, let us ignore effects of
$P(\tau,t)$, and consider the Fourier-transform $K(w)$ of $
K(t)=\int dw K(w)e^{-i(w-\omega_L)t}$. From Eq. (\ref{kappa1}),
one obtains
\begin{eqnarray}
\kappa(t)= \int\limits_{t_0}^t d\tau \int
dwK(w)e^{-i(w-\omega_L)(\tau-t)}D_{++}(\tau-t) \label{kappa3}
\end{eqnarray}
for the dephasing rate. Notice that the Markovian approximation
holds whenever $K(\omega)$ is smooth in the vicinity of both the
frequency $\omega_L$ and  TLS transition frequency. Moreover, Eq.
(\ref{kappa3}) indicates that a sufficient intense driving-field
probes $K(\omega)$ away from the $\omega_L$ frequency. The
$K(\omega)$ spectrum may be smooth enough in the vicinity of all
components of the triplet $\omega_L, \omega_L \, \pm \, \Omega_{R}$
to justify a Markovian approximation for each of them
\cite{{apn-works},{florescu}}. Taking into consideration that
$K(\omega)$ has different values at these frequencies, even a
Markovian approximation for each component of the triplet should
yield to an intensity-dependent dephasing rate. Therefore, by
performing the Markovian approximation for the different components
of the triplet in a standard way, for a rectangular driving pulse,
one finds from Eq. (\ref{kappa3}) the time-independent dephasing
rate \cite{apn-works}: $\kappa\approx
\frac{\pi}{2}(c^2+1)K(\omega_L)+
\frac{\pi}{4}s^2[K(\omega_L+\Omega_{R})+ K(\omega_L-\Omega_{R})]$.
Moreover, when differences in values of $K(\omega)$ at frequencies
$\omega_L$, $\omega_L \, \pm \, \Omega_{R}$ are much smaller than
the value of $K(\omega_L)$, one may expand $K(\omega)$ in the
vicinity of $\omega_L$ and obtain
\begin{eqnarray}
\kappa=\pi K(\omega_L)+ {\pi\Omega^2\over 4}{d^2\over
d\omega^2}K(\omega)\Bigr|_{\omega=\omega_L}\label{kappa5}
\end{eqnarray}
as an intensity-dependent dephasing rate. Here we notice that Brandi
\textit{et al} \cite{oliveira} have used an intensity-dependent
recombination rate as in Eq. (\ref{kappa5}) to model experimental
measurements on ROs in a QD semiconductor TLS, and found good
agreement with the excitonic photocurrent data as measured by
Zrenner \textit{et al} \cite{exper}. Also, from Eq. (\ref{kappa2}),
one may use the same approximation as before in obtaining Eq.
(\ref{kappa5}), and find
\begin{eqnarray}
\overline{\Omega}=\frac{1}{2}\left\{\Omega- \pi\Omega {d\over
d\omega}K(\omega)\Bigr|_{\omega_L}+ {\pi \Omega \Delta
\over2}{d^2\over d\omega^2}K(\omega)\Bigr|_{\omega_L}\right\}\label{rabi3}
\end{eqnarray}
for the generalized time-independent Rabi frequency.

Now we apply the developed approach in order to obtain a
quantitative understanding of the experimental measurements by
Zrenner \textit{et al} \cite{exper} and Wang \textit{et al}
\cite{no-biexciton}. Figure 1 displays the present results
corresponding to the solution of the Bloch equations with the
driving-dependent dephasing rate and generalized Rabi frequency
[see Eqs. (\ref{kappa5}) and (\ref{rabi3})] chosen in order to
give the appropriate ROs as found in the experimental measurements
\cite{exper,no-biexciton}. One clearly notes the excellent
agreement with the excitonic photocurrent measurements of Zrenner
\textit{et al} \cite{exper} [Fig. 1(a)] and photoluminescence
measurements by Wang \textit{et al} \cite{no-biexciton} [Fig.
1(b)]. One needs to emphasize that, with respect to the effects
stemming from the stationary part of the reservoir's correlation
function, the particular form of the $K(\omega)$ function is of no
importance as long as it satisfies quite general requirements as
mentioned before. In the present calculation only the value of the
$K(\omega)$ function at the point $\omega_L$ and two of its
derivatives are of importance [cf. Eqs. (\ref{kappa5}) and
(\ref{rabi3})]. These are the only "free" parameters to match the
experiment. Moreover, apart from the value $K(\omega_L)$, only the
second derivative of $K(\omega)$ at the point $\omega_L$ plays a
significant role, and we have actually used essentially this
parameter to match the experimental data.

\begin{figure}
\epsfig{figure=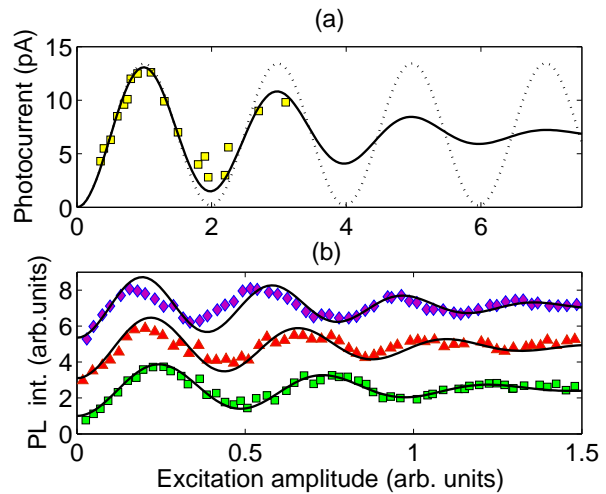,width=\linewidth}
\caption{(a) Rabi oscillations of the photocurrent, at resonance, as
a function of the excitation amplitude. The dotted line is the
solution given by the Markovian Bloch equations with the dephasing
rate independent of the driving field, whereas the solid line
corresponds to the solution of the Bloch equations with the
driving-dependent dephasing rate and generalized Rabi frequency
given by Eqs. (\ref{kappa5}) and (\ref{rabi3}). Full squares
represent experimental data from Zrenner \textit{et al}
\cite{exper}, for a pulse width of about 1 ps. Here, a $\pi$-pulse
corresponds to the unit of the excitation amplitude;
(b) ROs in the photoluminescence (PL) intensity, at resonance,
with full theoretical curves corresponding in descending order to
pulse widths of 9.3 ps, 7.0 ps, and 5.4 ps, respectively.
Calculations are performed with the driving-dependent dephasing rate
and generalized Rabi frequency as in (a). Full symbols are the
corresponding experimental data from Wang \textit{et al}
\cite{no-biexciton}.} \label{fig1}
\end{figure}

\begin{figure}
\epsfig{figure=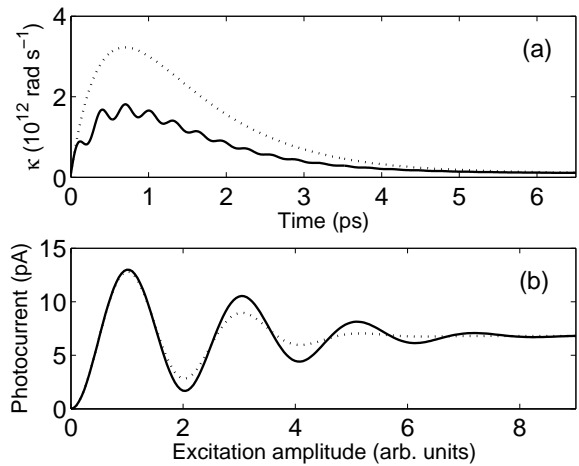,width=\linewidth}
\caption{Examples of (a) dephasing rate $\kappa(t)$ [cf. Eq.
(\ref{kappa6})], and (b) upper-state population dynamics, for the
fixed time moment corresponding to the end of the rectangular pulse,
versus excitation amplitude, for a model \cite{apn-works}
$\langle{\bf R}(t)\rangle\sim\gamma\Omega e^{-\gamma(t-t_0)}$, with
$\gamma = 2 \,ps^{-1}$. Dotted lines correspond to $D_{++}(\tau-t) =
1$ whereas solid lines correspond to the $D_{++}(\tau-t)$ defined
in Eq. (\ref{dfunctions1}). } \label{fig3}
\end{figure}

We now consider that the coherent driving-pulse applied to the TLS
may also influence its surroundings.  If the action of the driving
field on the system surroundings is weak, the $K(\tau-t)$ stationary
contribution to the reservoir will essentially have the same
dependence on the driving-field intensity as described above. We
note that the driving-pulse action on the reservoir may also give
rise to non-Markovian effects stemming from the $P(\tau,t)$
non-stationary part of the reservoir's correlation function, and
that observable manifestations of these effects may be very similar
to those described above. Let us illustrate it with a simple model
of a bosonic reservoir driven by the same rectangular pulse that is
applied on the TLS under investigation. Using the RWA, one may
describe the whole "TLS + reservoir" system with the following
Hamiltonian
\begin{eqnarray}
\nonumber H_{1}(t)&=& H_0(t)+ H_{res}(t) \\
&+& \hbar\sigma^+\sigma^-\sum\limits_j [g_j \,
b_j^+e^{i\omega_L(t-t_0)}+h.c.], \label{ham3}
\end{eqnarray}
where $H_{res}(t)$ is the reservoir Hamiltonian,
\begin{equation}
H_{res}(t)=\hbar\sum\limits_j\Delta_jb_j^+b_j+\sum\limits_j\Omega_j(t)(b_j^++b_j),
\end{equation}
and the $g_j$ are interaction constants, the $\Delta_j$ are
detunings of the reservoir modes from the driving field, and the
$\Omega_j$ are Rabi frequencies for every particular reservoir mode
(we assume them to be constant, $\Omega_j(t)\equiv\Omega_j$, for the pulse duration).
Using the interaction picture with respect to $H_{res}(t)$, one recovers from Eq. (\ref{ham3})
the Hamiltonian of Eq. (\ref{ham1})  with the following reservoir
operator
\begin{eqnarray}
{\bf
R}(t)=\sum\limits_jg_j\,[b_j+\frac{\Omega_j}{\Delta_j}]\,e^{i(\Omega_L+\Delta_j)(t_0-t)}+h.c.
\, , \label{r1}
\end{eqnarray}
and with the system's detuning  shifted due to the interaction with
the excited reservoir, i.e.,
$\Delta\rightarrow\Delta-2\sum\limits_jg_j{\Omega_j/\Delta_j}$.  For
the reservoir initially at the vacuum state, one obtains
\begin{eqnarray}
\langle{\bf
R}(t)\rangle=\sum\limits_jg_j{\Omega_j\over\Delta_j}e^{i(\Omega_L+\Delta_j)(t_0-t)}+h.c.
\,, \label{avr1}
\end{eqnarray}
and the reservoir correlation function $\langle{\bf R}(\tau){\bf
R}(t)\rangle$ as the sum of a stationary part
$K(\tau-t)=\sum\limits_jg_j^2e^{i(\Omega_L+\Delta_j)(t-\tau)}$
with a non-stationary part $P(\tau,t)=\langle{\bf
R}(\tau)\rangle\langle{\bf R}(t)\rangle$. The stationary part
$K(\tau-t)$ produces effects already described above, and here we
assume that $K(w)$ is wide and smooth enough so that the
stationary dephasing rate, $\kappa_s$ , is independent of the
intensity of the driving field. Then, from Eq. (\ref{kappa1}), one
has
\begin{eqnarray}
\kappa(t)=\kappa_s+\langle{\bf R}(t)\rangle\int\limits_{t_0}^t d\tau
\langle{\bf R}(\tau)\rangle D_{++}(\tau-t).\label{kappa6}
\end{eqnarray}
Note that the non-stationary part of the dephasing rate $\kappa(t)$
decays with time [see Fig.~\ref{fig3}(a)], since $\langle{\bf
R}(t)\rangle\rightarrow0$ for $t\rightarrow \infty$, and that the
function $\langle{\bf R}(t)\rangle$ may decay slower than the
stationary part, $K(t)$, of the reservoir's correlation function as
the driving field excites different modes of the reservoir in a
different way, and the spectral density of the reservoir's
excitation may therefore be much narrower than $K(w)$. Moreover, in
experiments on ROs in localized semiconductor systems one deals with
short driving pulses, so that the non-stationary part of the
dephasing rate may play a significant part in the system's dynamics.
Even if one assumes $D_{++}(t)\approx 1$ for the time-interval of
interest, the non-stationary part of the dephasing rate will be
dependent on the driving-field intensity. This is a purely
non-Markovian dynamical effect producing an intensity-dependent
damping of  ROs [cf. Fig.~\ref{fig3}(b)] quite similar to those
described before.

The decrease of the dephasing rate with time may be responsible
for the constant value of the dephasing rate as measured
\textit{after} the application of the driving pulse in the
experimental measurements by Patton \textit{et al}
\cite{int_indep} [this situation is illustrated in
Fig.~\ref{fig3}(a)]. Also, it may explain the decreased dephasing
rate after the application of the pulse as in the experiment by
Wang \textit{et al} \cite{no-biexciton}. To conclude, a
driving-dependent damping of ROs due to the non-stationary
contribution of the reservoir's correlation function may take
place for quite general reservoirs. Indeed, the nature of the
reservoir influences only the particular form of $P(\tau,t)$ and
not its general properties, which determine the effect in
question.

Summing up, we have demonstrated that the damping of ROs with the
driving-field intensity in localized semiconductor systems (QDs,
shallow donors, etc) is an effect of a very general nature, and a
consequence of non-Markovian effects due to the coupling of the
system to a reservoir. The exact nature of a reservoir (being an
ensemble of phonons, other localized systems, traps, free carriers
in a wetting layer, coupling to bi-excitons or higher decaying
levels, etc, or a combination of mechanisms) is not of particular
importance for the manifestation of the effect. Similar damping of
ROs may occur as a consequence of different physical mechanisms. The
first one stems from stationary properties of the reservoir whereas
the second one is a purely non-stationary effect occurring when the
driving field excites the reservoir with a decay time of the
non-stationary part of the reservoir's correlation function
comparable to the driving-field pulse length.

The authors gratefully acknowledge partial financial support by EU
under EQUIND project of 6FP IST-034368 and INTAS, and by Brazilian
Agencies CNPq, FAPESP, Rede Nacional de Materiais
Nanoestruturados/CNPq, MCT - Millenium Institute for Quantum
Information, and  MCT - Millenium Institute for Nanotechnology.


\begin{thebibliography}{99}


\bibitem{experiment} B. E. Cole, J. B. Williams, B. T. King, M .S.
Sherwin, and C. R. Stanley, Nature (London) \textbf{410}, 60 (2001).

\bibitem{exper} A. Zrenner, E. Beham, S. Stufler, F. Findels,
M. Bichler, G. Abstreiter, Nature \textbf{418}, 612 (2002).


\bibitem{no-biexciton} Q. Q. Wang, A. Muller, P. Bianucci, E. Rossi,
Q. K. Xue, T. Takagahara, C. Piermarocchi, A. H. MacDonald, and C. K.
Shih, Phys. Rev. B \textbf{72}, 035306 (2005).


\bibitem{int_indep}B. Patton, U. Woggon, and W. Langbein, Phys. Rev.
Lett. \textbf{95}, 266401 (2005).


\bibitem{traps} A. Berthelot, I. Favero, G. Cassabois, C. Voisin, C.
Delalande, Ph. Roussignol, R. Ferreira and J. M. Gerrard, Nature
Physics  \textbf{2}, $N^o$ 11, 758 (2006).

\bibitem{apn-works} S. Ya. Kilin and A. P. Nizovtsev, J. Phys. B:
At. Mol. Phys. \textbf{19}, 3457 (1986); P. A. Apanasevich, S. Ya.
Kilin, A. P. Nizovtsev, and N. S. Onischenko, J. Opt. Soc. Am. B
\textbf{3}, 587 (1986); J. Appl. Spectr. \textbf{47}, 1213 (1987).

\bibitem{spin} For a recent review, see
Y. Tanimura, J. Phys. Soc. Jap. \textbf{75}, 082001 (2006).



\bibitem{phonon} J. F\"{o}rstner, C. Weber, J. Danckwerts, and A. Knorr,
Phys. Rev. Lett. \textbf{91}, 127401 (2003).

\bibitem{deloc} A. Vasanelli, R. Ferreira, G. Bastard, Phys. Rev.
Lett. \textbf{89}, 216804 (2002).

\bibitem{biexciton} L. Besombes, J. J. Baumberg, and J. Motohisa,
Semicond. Sci. Technol. \textbf{19}, 148 (2004); J. M. Villas-Boas,
S. E. Ulloa, and A. O. Govorov, Phys. Rev. Lett. \textbf{94}, 057404
(2005).

\bibitem{oliveira} H. S. Brandi, A. Latg\'{e}, and  L. E. Oliveira,
Phys. Rev. B \textbf{68}, 233206 (2003); H. S. Brandi, A. Latg\'{e},
Z. Barticevic, L. E. Oliveira, Solid State Commun. \textbf{135}, 386
(2005).

\bibitem{shih1} P. Bianucci, A. Muller, and C. K. Shih, Q. Q. Wang and Q. K.
Xue, C. Piermarocchi, Phys. Rev. B \textbf{69}, 161303(R) (2004).


\bibitem{science} X. Li, Y. Wu, D. Steel, D. Gammon, T.H. Stievater, D.S. Katzer,
D. Park, C. Piermarocchi, and L.J. Sham, Science \textbf{301}, 809
(2003).

\bibitem{tlmast} A. Royer, Phys. Rev. A \textbf{6}, 1741 (1972);
H.-P. Breuer, B. Kappler and F. Petruccione Phys. Rev. A
\textbf{59}, 1633 (1999).

\bibitem{dress} R. R. Puri, \textit{Mathematical Methods of
Quantum Optics}, Berlin: Springer, 2001.

\bibitem{florescu} M. Florescu and S. John, Phys. Rev. A
\textbf{69}, 053810 (2004).

\end{thebibliography}
\end{document}